# Elastic and electro-optical properties of flexible fluorinated dimers with negative dielectric anisotropy

Greta Babakhanova[a,#], Hao Wang[a,b], Mojtaba Rajabi[a,c], David Li[a],

Quan Li[a,b], and Oleg D. Lavrentovich[a,b,c]

[a]*Advanced Materials and Liquid Crystal Institute, Kent State University, Kent, USA*

[b]*Materials Science Graduate Program, Kent State University, Kent, USA*

[c]*Department of Physics, Kent State University, Kent, USA*

[#]Present address: National Institute of Standards and Technology, Gaithersburg, MD, USA.

Oleg D. Lavrentovich: olavrent@kent.edu (Corresponding author)

**Abstract**

We report the synthesis and the temperature dependencies of optical, dielectric, and elastic properties of four homologous dimeric mesogens in the uniaxial nematic (N) phase. The studied dimers are derived from 2',3'-difluoroterphenyls units connected via flexible alkyl chains. The compounds differ in the length of either the flexible alkyl chain or the terminal groups. These achiral bimesogens exhibit the twist-bend nematic ($N_{TB}$) phase with a nanoscale pitch of the modulated nematic director, as established by the freeze-fracture transmission electron microscopy. All four homologs have negative dielectric anisotropy owing to the permanent dipole oriented normal to the long molecular axes. A longer flexible bridge length yields an expanded temperature range of the N phase, stronger dielectric anisotropy, larger birefringence and bend elastic constant $K_{33}$. Longer alkyl terminal chains increase birefringence but decrease the absolute value of dielectric anisotropy and $K_{33}$. All four bimesogens show a dramatic decrease of $K_{33}$ down to ~0.5 pN on approaching the N-$N_{TB}$ transition. A longer bridge yields a longer period of twist-bend modulation in the $N_{TB}$ phase.

## 1. Introduction

The material properties of liquid crystals depend strongly on the shape of their molecules. The simplest rod-like molecules produce a uniaxial nematic (N), widely used in applications. Lately, there has been much interest in the so-called bent-core molecules, formed by two rod-like segments attached to each other at some angle. This kinked shape leads to fundamentally new properties and phases, as reviewed recently [1, 2]. For example, bent-core molecules synthesized by B.K. Sadashiva and his group, develop a biaxial smectic A phase either by themselves [3, 4, 5] or when embedded into an otherwise uniaxial smectic [6, 7]. Nematics formed by molecules with an obtuse bent angle often exhibit a bend elastic constant that is smaller than the splay constant, $K_{33} < K_{11}$, [1, 2, 8, 9, 10, 11, 12, 13], which is opposite to the trend $K_{33} > K_{11}$ found in conventional rod-like nematics [14], acute-angle bent-core nematics [15] and mixtures of the two types [16]. The tendency to bend might be so strong that the bent-core materials exhibit new states with spatially-varying orientation, such as the twist-bend nematic $N_{TB}$ [17, 18, 19, 20, 21] or an oblique helicoidal cholesteric [22, 23].

The nanoscale structure of $N_{TB}$ with a periodic twist and bend of molecular orientations is clearly seen in transmission electron microscopy studies performed by two independent research groups [20, 21, 24]. The $N_{TB}$ phase occurs when the mesogenic molecules favor bend conformations. Molecules with two rigid cores connected by a flexible aliphatic chain with an odd number of methylene groups turned out to be conducive to form $N_{TB}$ phase since the prevailing conformation of the flexible bridge produces an obtuse angle between two rod-like rigid units. The tendency of molecules to bend was described by Dozov as a negative value of the nematic bend modulus $K_{33}$ [18]. To assure a constant bend curvature in space, the bend must be accompanied by a twist [17, 18, 25]. As a result, $N_{TB}$ molecular orientation follows an oblique helicoid with a pitch on the order of 10 nm [20, 21, 24, 26, 27, 28, 29, 30, 31]. The structure is stable provided the twist elastic constant $K_{22}$ is significantly smaller than its splay counterpart $K_{11}$, satisfying the condition $K_{11}/K_{22} > 2$ [17, 18, 32]. Conventional rod-like calamitic liquid crystals do not satisfy this condition, as the observed trend is $K_{33} > K_{11} > K_{22}$ [33]. In a parallel vein, there was a recognition that the disparity of elastic constants needed for the existence of N$_{TB}$ is also involved in guaranteeing the stability of a peculiar chiral nematic state, the oblique helicoidal cholesteric, abbreviated Ch$_{OH}$, which exists when either an electric [22, 34] or magnetic field [35] is applied. This state, predicted by Meyer [25] and de Gennes [36], has been realized in flexible dimeric nematic materials with a small $K_{33}$ [22, 34]. The observed connection between the types of molecular packings and elasticity has attracted a strong research interest to the measurements of elastic constants of flexible dimers in the uniaxial nematic (N) [13, 19, 37, 38, 39, 40, 41, 42] and in the cholesteric (Ch) phase [23]. It was found that $K_{33}$ in N and Ch formed by flexible dimers can reach anomalously low values, about 0.5 N near the N-N$_{TB}$ transition temperature, $T_{NTB}$, [13, 19, 23, 37, 39, 41, 42, 43, 44, 45], or the analog of this transition in chiral materials [23]. Other physical properties of the flexible dimers also deviate from the behavior of rigid rod-like mesogens. For instance, in some dimers, the temperature dependencies of birefringence, $\Delta n(T)$, and dielectric anisotropy, $\Delta \varepsilon(T)$, are nonmonotonous [13, 21, 41, 45, 46].

The structure of the spacer greatly affects the phase behavior of dimers [47, 48, 49, 50]. An obvious example is that the dimers with an even number of methylene groups in the flexible link do not produce the N$_{TB}$ phase, whereas their odd-numbered counterparts do. Recently, Imrie, Mehl, Goodby, Mandle, and their colleagues presented extensive studies of the relationship between the chemical structures of flexible dimers and the occurrence of the N$_{TB}$ phase [51, 52, 53, 54, 55, 56, 57, 58]. This work complements these studies with the exploration of the physical properties of the uniaxial N phase formed by flexible dimers that produce a negative dielectric anisotropy $\Delta \varepsilon = \varepsilon_{||} - \varepsilon_{\perp}$, where the subscripts indicate the orientation of the probing electric field with respect to the director. The goal is to establish how the chemical structure affects the temperature dependences of $K_{33}$, $\Delta\varepsilon$, and birefringence $\Delta n = n_{\parallel} - n_{\perp}$ in the group of four compounds with a common pair of rigid rod-like 2',3'-difluoroterphenyls units connected by a flexible alkyl chain, Fig.1. The homologs differ in the linking chain length as well as the length of the terminal moiety. Freeze-fracture electron microscopy (FFTEM) confirms the existence of the $N_{TB}$ phase with a nanoscale heliconical structure in all four compounds. Longer bridging links expand the temperature range of the N phase and increase $\Delta n(T)$, $|\Delta\varepsilon(T)|$ as well as $K_{33}$. Increasing the terminal units from $C_3H_7$ to $C_5H_{11}$ increases $\Delta n(T)$ but decreases $|\Delta\varepsilon(T)|$ and $K_{33}$. All four homologs exhibit extremely low values of $K_{33}$ as one approaches $T_{NTB}$. We first present the synthezis of the materials in Section 2 and report on their properties in Section 3.

## 2. Synthesis of Dimeric Mesogens

The synthesis of the fluorinated dimeric mesogens presented in Fig. 1 is based on techniques described in Refs. [20, 21, 59, 60]. In order to optimize the synthesis route, we do not follow one particular approach, but combine different protocols, as detailed below.

All reagents and solvents were commercially available and used as received unless otherwise stated. Chemicals (2,3-difluoro-4'-pentyl[1,1'-biphenyl]-4-yl)boronic acid and (2,3-difluoro-4'-propyl[1,1'-biphenyl]-4-yl)boronic acid were purchased from Kingston Chemicals Ltd. The rest of the chemical materials were acquired from Sigma-Aldrich and TCI America. [1]H (400 MHz), [13]C (100 MHz) and [19]F (376 MHz) NMR spectra were recorded on a Bruker Nuclear Magnetic Resonance (NMR) spectrometer using CDCl$_3$ as a solvent, see Supplement. Chemical shifts are in δ unit (ppm) with the residual solvent peak as the internal standard. The coupling constant ($J$) is reported in Hertz (Hz). NMR splitting patterns are designated as follows: s, singlet; d, doublet; t, triplet; and m, multiplet. Column chromatography is carried out on silica gel (230-400 mesh). Analytical thin-layer chromatography (TLC) is performed on commercially coated 60 mesh $F_{254}$ glass plates. Spots on the TLC plates are rendered visible by exposure to UV light.

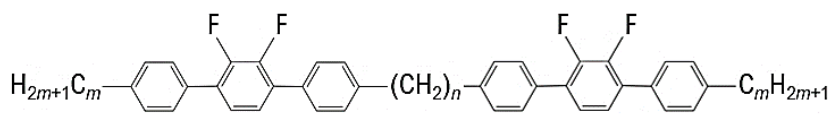

**Figure 1.** The chemical structures of DTC5C7 (*m*=5, *n*=7), DCT5C9 (*m*=5, *n*=9), DTC5C11 (*m*=5, *n*=11) and DTC3C11 (*m*=3, *n*=11).

### 2.1 Synthetic route of DTC5C7

Bromobenzene (30 mL, 287 mmol) is added to the flask of aluminum chloride (7.07 g, 53 mol) and rapidly stirred under nitrogen for 1 hour (Fig. 2a). After the solution is cooled in an ice-water bath, pimeloyl dichloride (5 g, 25 mmol) is dissolved in bromobenzene (5 mL, 48 mmol) and is added dropwise to the previous flask for over one hour. The solution is then heated to 45 °C and stirred for 48 hours. The reaction is cooled to ambient temperature and poured into a solution of concentrated hydrochloric acid (100 mL) in an ice-water bath and water (100 mL) is added. Subsequently, dichloromethane is added to the solution to extract the organic compound four times. The organic part is dried over magnesium sulfate overnight. Dichloromethane is evaporated with a rotatory evaporator. The crude solid is purified by silica gel chromatography column with the eluent of hexane: ethyl acetate (1:1 volume ratio) to get 1,7-bis(4-bromophenyl)heptane-1,7-dione as colorless crystal with yield 10.26 g, 92 %. [1]H NMR (400 MHz, CDCl₃): δ 7.81 (d, *J* = 8.6 Hz, 4H), 7.59 (d, *J* = 8.6 Hz, 4H), 2.92 (t, *J* = 7.3 Hz, 4H), 1.76-1.69 (m, 4H), 1.42-1.35 (m, 2H). [13]C NMR (100 MHz, CDCl₃): δ 199.10, 135.69, 131.89, 129.56, 128.10, 38.22, 28.82, 23.88.

1,7-bis(4-bromophenyl)heptane-1,7-dione (5.53 g, 12.6 mmol) is dissolved in trifluoroacetic acid (20 mL, 29.8 g, 0.26 mol), Fig. 2b. Triethylsilane (10 mL, 7.3 g, 0.06 mol) is added dropwise over one hour with rapid stirring in an ice-water bath. The reaction is stirred at ambient temperature for 48 hours. Then the reaction mixture is poured into an ice-water mixture. The organic compound is extracted by hexane four times and dried over magnesium sulfate overnight. The solvent is then evaporated and the crude product is separated by column (silica gel) chromatography with an eluent of hexane/dichloromethane 2/1 and 1,7-bis(4-bromophenyl)heptane is acquired with a yield of 4.66 g, 90%. [1]H NMR (400 MHz, CDCl$_3$): δ 7.38 (d, $J$ = 8.4 Hz, 4H), 7.04 (d, $J$ = 8.6 Hz, 4H), 2.55 (t, $J$ = 7.7 Hz, 4H), 1.61-1.53 (m, 4H), 1.35-1.24 (m, 6H). [13]C NMR (100 MHz, CDCl$_3$): δ 141.69, 131.24, 130.13, 119.25, 35.28, 31.22, 29.22, 29.00.

A solution of 1,7-bis(4-bromophenyl)heptane (1 g, 2.44 mmol) and (2,3-difluoro-4'-pentyl-[1,1'-biphenyl]-4-yl)boronic acid (1.85 g, 6.1 mmol) in a mixture of 1,2-dimethyloxyethane (30 mL) and saturated aqueous solution of sodium carbonate (20 mL) is thoroughly degasses and refilled with nitrogen three times, after which, Tetrakis(triphenylphosphine)palladium (50 mg) is added, Fig. 2c. The reaction is refluxed at 125 °C for 48 hours under stirring. The reaction is then cooled to ambient temperature and water is added to the reaction. The aqueous solution is extracted with dichloromethane four times and dried over magnesium sulfate overnight. The organic solvent is evaporated and the crude product is separated by column chromatography (eluent of hexane/dichloromethane: 2/1 by volume) and yields 1,7-bis (2', 3'-difluoro-4''-pentyl-[1,1':4',1''-terphenyl]-4-yl)heptane, DTC5C7, (Fig. 2d) 1.44 g, 77%. [1]H NMR (400 MHz, CDCl$_3$): δ 7.50 (d, $J$ = 8.0 Hz, 8H), 7.28 (d, $J$ = 8.0 Hz, 4H), 7.22 (m, 4H), 2.64 (t, $J$ = 8.0 Hz, 8H), 1.70-1.62 (m, 8H), 1.39-1.38 (m, 14H), 0.91 (t, $J$ = 7.1 Hz, 6H). [13]C NMR (100 MHz, CDCl$_3$): δ 149.79, 149.64, 147.30, 147.15, 143.07, 143.98, 131.99, 131.96, 129.47, 128.66, 124.54, 35.68, 31.56, 31.33, 31.10, 29.32, 29.20, 22.56, 14.04. [19]F NMR: δ -143.31, -143.32.

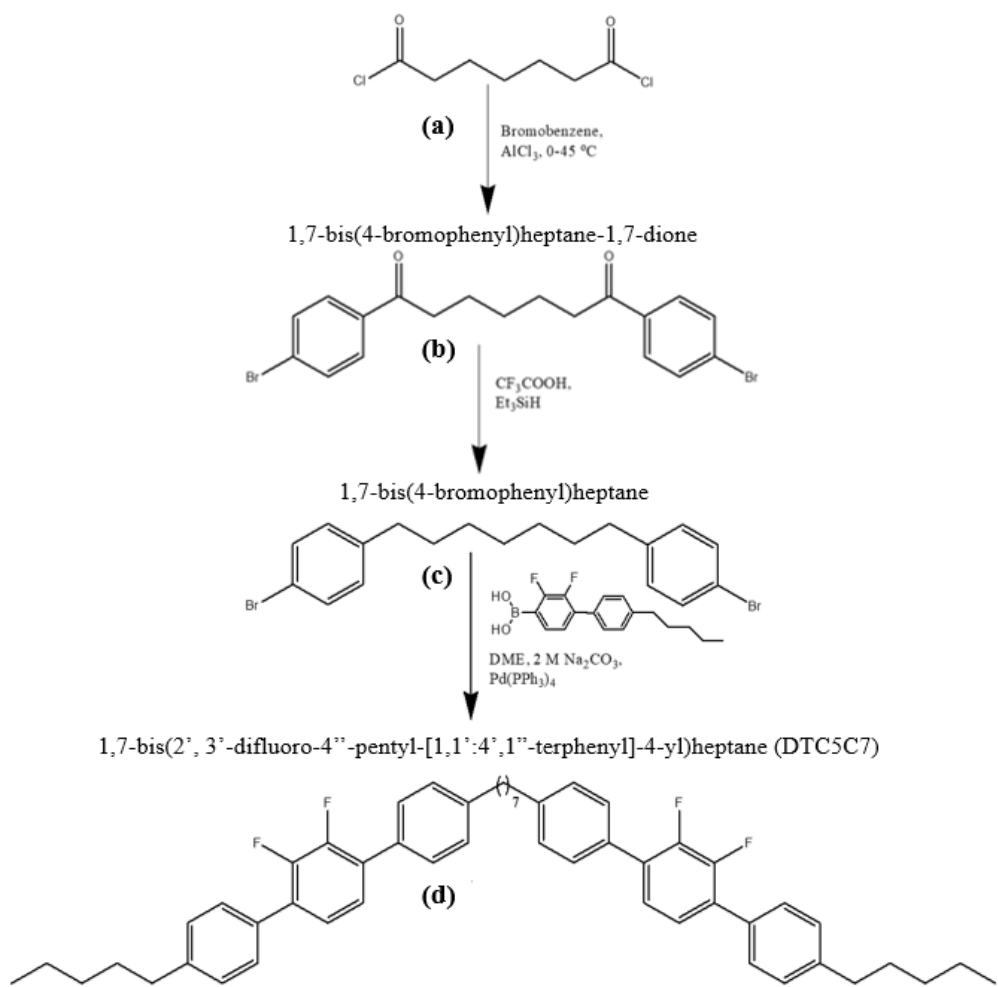

**Figure 2.** Synthetic scheme of DTC5C7.

*2.2. Synthetic route of DTC5C9*

The synthesis of compounds in Fig. 3a,b is similar to the process outlined in Fig. 2a,b and *Section 2.1*. The results for 1,9-bis(4-bromophenyl)nonane-1,9-dione are: [1]H NMR (400 MHz, CDCl₃): δ 7.81 (d, *J* = 8.6 Hz, 4H), 7.59 (d, *J* = 8.6 Hz, 4H), 2.92 (t, *J* = 7.3 Hz, 4H), 1.76-1.69 (m, 4H), 1.42-1.35 (m, 6H). [13]C NMR (100 MHz, CDCl₃): δ 199.37, 135.72, 131.86, 129.58, 128.03, 38.48, 29.25, 29.08, 24.10; for 1,9-bis(4-bromophenyl)nonane: [1]H NMR (400 MHz, CDCl₃): δ 7.38 (d, *J* = 8.4 Hz, 4H), 7.04 (d, *J* = 8.6 Hz, 4H), 2.55 (t, *J* = 7.7 Hz, 4H), 1.61-1.53 (m, 4H), 1.35-1.24 (m, 10H). [13]C NMR (100 MHz, CDCl₃): δ 141.77, 131.21, 130.14, 119.21, 35.32, 31.28, 29.40, 29.12.

The synthesis of 1,9-bis(2',3'-difluoro-4''-pentyl-[1,1':4',1''-terphenyl]-4-yl)nonane, DTC5C9, (Fig. 3c) is similar process shown in Fig. 2c in *Section 2.1*. The yeild of DTC5C9 is 0.9 g (Fig. 3d). $^1$H NMR (400 MHz, CDCl$_3$): δ 7.50 (d, $J$ = 8.0 Hz, 8H), 7.28 (d, $J$ = 8.0 Hz, 4H), 7.22 (m, 4H), 2.66 (t, $J$ = 8.0 Hz, 8H), 1.70-1.62 (m, 8H), 1.39-1.38 (m, 18H), 0.91 (t, $J$ = 7.1 Hz, 6H). $^{13}$C NMR (100 MHz, CDCl$_3$): δ 149.79, 149.63, 147.30, 147.14, 143.06, 143.04, 131.96, 129.45, 128.69, 128.66, 124.57, 125.54, 35.69, 31.56, 31.39, 31.10, 29.46, 29.32, 22.56, 14.04.

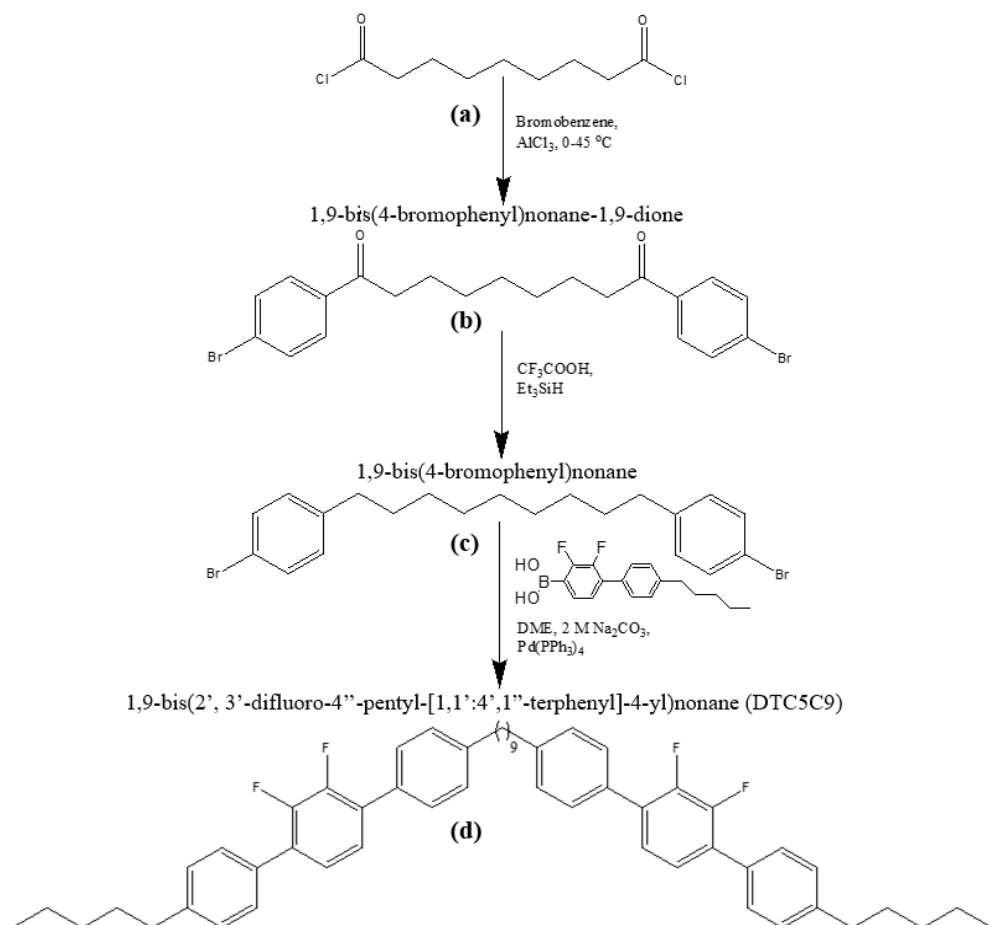

**Figure 3**. Synthetic scheme of DTC5C9.

### 2.3. Synthetic route of DTC5C11

Undecanedioic acid (1 g, 4.6 mmol) is dissolved in dry dichloromethane (5 mL) under nitrogen. Oxalyl chloride (1.6 mL, 2.34 g, 18.4 mmol) is added followed by dry N,N'-dimethylformamide (1 drop), and gentle stirring (Fig. 4a). The reaction is stirred overnight and then refluxed for 2 hours until gas evolution is ceased. The solution is evaporated by vacuum distillation to yield crude undecanedioyl dichloride which is used immediately in the Friedel-Crafts acylation described below.

Bromobenzene (5.8 mL, 8.7 g, 55.4 mmol) is added to the powdered aluminum chloride (1.3 g, 9.8 mmol) and rapidly stirred under nitrogen for 1 hour, Fig. 4b. The crude undecanedioyl chloride is dissolved in bromobenzene (1.3 mL, 1.95 g, 12.4 mmol) and is added dropwise over 1 hour in an iced water bath. The reaction is then heated to 40 °C and stirred for 48 hours. Afterward, the reaction is cooled to ambient temperature and poured into a solution of concentrated hydrochloric acid (100 mL) in an ice-water bath. The aqueous layer is extracted with dichloromethane four times and the organic extracts are dried over magnesium sulfate with stirring overnight. The dichloromethane is evaporated and the crude solid is purified by silica gel column chromatography to yield the colorless crystal 0.92 g, 40%. [1]H NMR (400 MHz, CDCl$_3$): δ 7.81 (d, $J$ = 8.6 Hz, 4H), 7.59 (d, $J$ = 8.6 Hz, 4H), 2.92 (t, $J$ = 7.3 Hz, 4H), 1.76-1.69 (m, 4H), 1.42-1.35 (m, 10H). [13]C NMR (100 MHz, CDCl$_3$): δ 199.46, 135.72, 131.84, 129.58, 127.99, 38.54, 29.36, 29.24, 24.19.

1,11-bis(4-bromophenyl)undecane-1,11-dione (0.92 g, 1.87 mmol) is dissolved in trifluoroacetic acid (3 mL, 4.5 g, 39.3 mmol), Fig. 4c. Triethylsilane (1.7 mL, 1.2 g, 10 mmol) is added dropwise for over one hour with rapid stirring in an ice-water bath. The reaction is stirred at ambient temperature for 48 hours. Then the reaction mixture is poured into an ice-water mixture. The organic compound is extracted by hexane four times and dried over magnesium sulfate overnight. The solvent is then evaporated and the crude product is separated by column (silica gel) chromatography with an eluent of hexane/dichloromethane 2/1 and yields 1,11-bis(4-bromophenyl)undecane, 0.78 g, 90 %. [1]H NMR (400 MHz, CDCl$_3$): δ 7.38 (d, $J$ = 8.4 Hz, 4H), 7.04 (d, $J$ = 8.6 Hz, 4H), 2.56 (t, $J$ = 7.7 Hz, 4H), 1.62-1.55 (m, 4H), 1.35-1.24 (m, 14H). [13]C NMR (100 MHz, CDCl$_3$): δ 141.78, 131.20, 130.13, 119.20, 35.32, 31.31, 29.57, 29.51, 29.43, 29.15.

A solution of 1,11-bis(4'-bromophenyl)undecane (0.2 g, 0.43 mmol) and (2,3-difluoro-4'-pentyl-[1,1'-biphenyl]-4-yl)boronic acid (0.3 g, 0.99 mmol) in a mixture of 1,2-dimethoxyethane (20 mL) and 2 M sodium carbonate solution (20 mL) is degassed with nitrogen three times, Fig. 4d. Tetrakis(triphenylphosphine)palladium (50 mg) is added. The reaction is refluxed at 125 °C for 24 hours under stirring. The reaction is then cooled to ambient temperature. Water is added to the reaction. The aqueous solution is extracted with dichloromethane four times and dried over magnesium sulfate overnight. The solvent is evaporated and the crude product is separated by column chromatography with an eluent of hexane/dichloromethane 1/1 and yields 1,11-bis(2',3'-difluoro-4"-pentyl-[1,1':4',1"-terphenyl]-4-yl)undecane, DTC5C11, (Fig. 4e), 288 mg, 81 %. $^1$H NMR (400 MHz, CDCl$_3$): δ 7.50 (d, $J$ = 8.0 Hz, 8H), 7.28 (d, $J$ = 8.0 Hz, 4H), 7.22 (m, 4H), 2.64 (t, $J$ = 8.0 Hz, 8H), 1.70-1.62 (m, 8H), 1.39-1.38 (m, 22H), 0.91 (t, $J$ = 7.1 Hz, 6H). $^{13}$C NMR (100 MHz, CDCl$_3$): δ 149.82, 147.33, 147.18, 143.07, 131.99, 129.48, 128.66, 124.54, 35.73, 35.69, 31.56, 31.39, 31.08, 29.62, 29.56 29.50, 29.36, 22.55, 14.03.

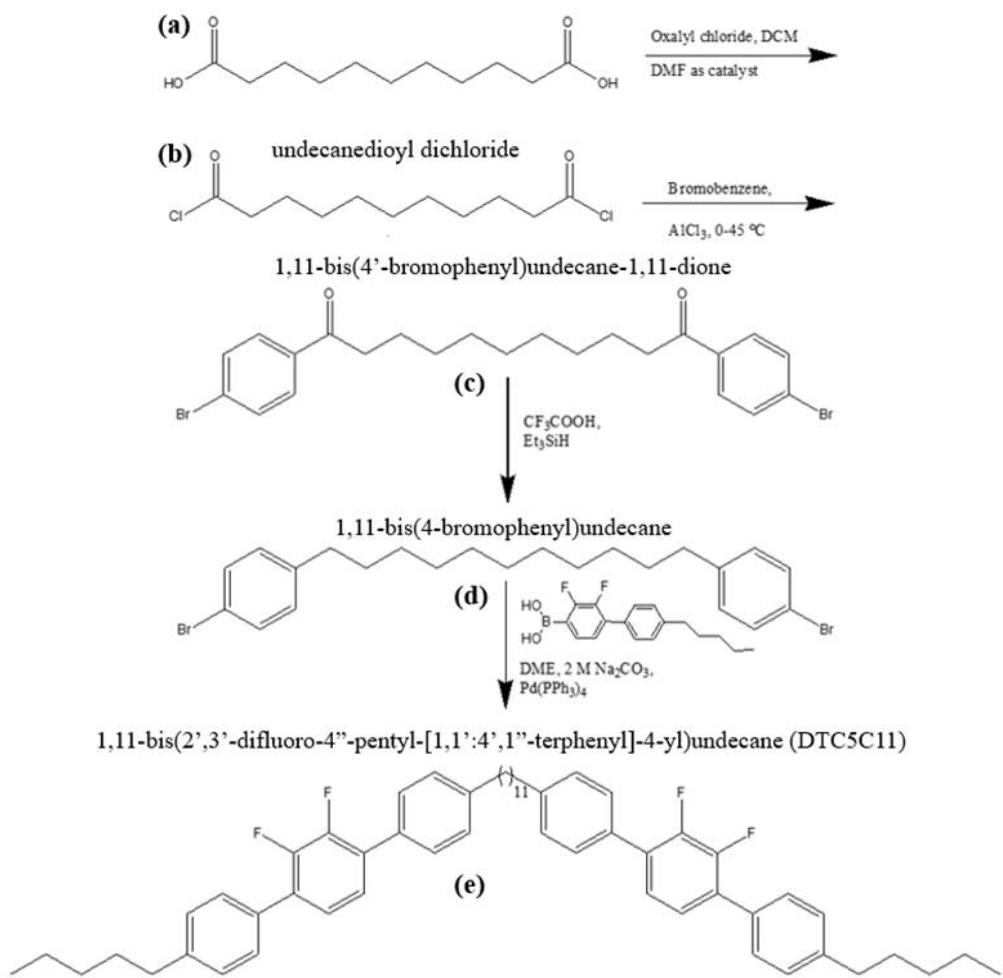

**Figure 4**. Synthetic scheme of DTC5C11.

### 2.4. Synthetic route of DTC3C11

A solution of 1,11-bis(4'-bromophenyl)undecane (0.2 g, 0.43 mmol) and (2,3-difluoro-4'-propyl-[1,1'-biphenyl]-4-yl)boronic acid (0.3 g, 1.09 mmol) in a mixture of 1,2-dimethoxyethane (20 mL) and 2 M sodium carbonate solution (20 mL) is degassed with nitrogen three times. Tetrakis(triphenylphosphine)palladium (50 mg) is added, Fig. 5a. The reaction is refluxed at 125 °C for 24 hours under stirring. The reaction is then cooled to ambient temperature. Water is added to the reaction. The aqueous solution is extracted with dichloromethane four times and dried over magnesium sulfate overnight. The solvent is evaporated and the crude product is separated by column chromatography with an eluent of hexane/dichloromethane 1/1 and yields 1,11-bis(2',3'-difluoro-4''-propyl-[1,1':4,1''-terphenyl]-4-yl)undecane, DTC3C11, (Fig. 5b), 250 mg, 76 %. [1]H NMR (400 MHz, CDCl$_3$): δ 7.50 (d, $J$ = 8.0 Hz, 8H), 7.28 (d, $J$ = 8.0 Hz, 4H), 7.22 (m, 4H), 2.64 (t, $J$ = 8.0 Hz, 8H), 1.70-1.62 (m, 8H), 1.39-1.38 (m, 14H), 0.91 (t, $J$ = 7.1 Hz, 6H). [13]C NMR (100 MHz, CDCl$_3$): δ 149.79, 149.63, 147.30, 147.14, 143.07, 142.80, 132.00, 131.96, 129.48, 128.71, 128.67, 124.54, 37.79, 35.72, 31.42, 29.62, 29.56, 29.50, 29.36, 24.48, 13.90.

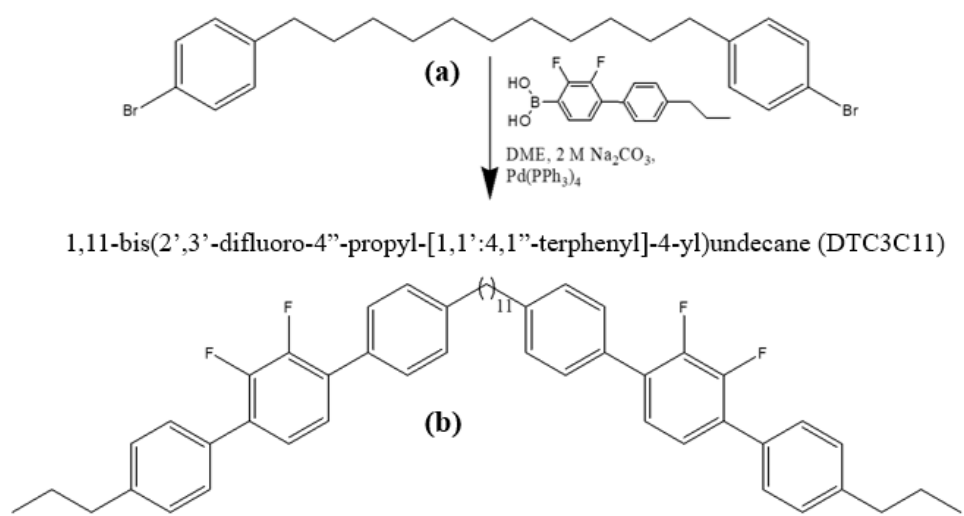

1,11-bis(2',3'-difluoro-4''-propyl-[1,1':4,1''-terphenyl]-4-yl)undecane (DTC3C11)

**Figure 5.** Synthetic scheme of DTC3C11.

## 3. Characterization of Material Properties

### 3.1 Molecular conformations and dipole moments

Usually, the molecular structure of the dimers based on 2',3'-difluoroterphenyl units are shown with the difluoro groups directed outward of the arched molecular shape, see Figs. 2a, 3d, 4e, 5b,12. The question is whether the actual molecular structure follows these schemes. For example, one might wonder whether the mutual orientation of dipoles associated with the difluoro groups might favor some well-defined subset of molecular conformations. To explore the issue, we perform numerical simulations of molecular conformations for DTC5C11, starting with three different initial states with a coplanar orientation of benzene rings and all difluoro groups. In conformation A, the two difluoro groups are directed outwards; in conformation B, one group is directed inward and another one outward; and in conformation C, both groups are directed inward. Next, molecular geometry is optimized by implementing density functional theory using B3LYP functional with 6-31G(d) basis set in Gaussian 09W software. The resulting energy-optimized bent conformations exhibit a net electric dipole $\mathbf{\mu}$ perpendicular to the long axis of the molecule, which correlates with the negative dielectric anisotropy of the materials. However, these optimized configurations exhibit different alignments of the difluoroterphenyl units and thus different absolute values of the dipoles, $\mu$ =3.83 D in the A case, 2.17 D for B, and 4.06 D for C. Since all optimized structures are of the same energy, the result suggests that the aromatic units of individual DTC5C11 molecules might adopt multiple orientations and that the most efficient conformations in the condensed $N_{TB}$ and N would be defined by interactions with neighboring molecules. Figure 6 presents the shapes and dipole moments for four homologs obtained by simulations that start with both difluoro groups pointing outwards but deviating slightly from a coplanar geometry.

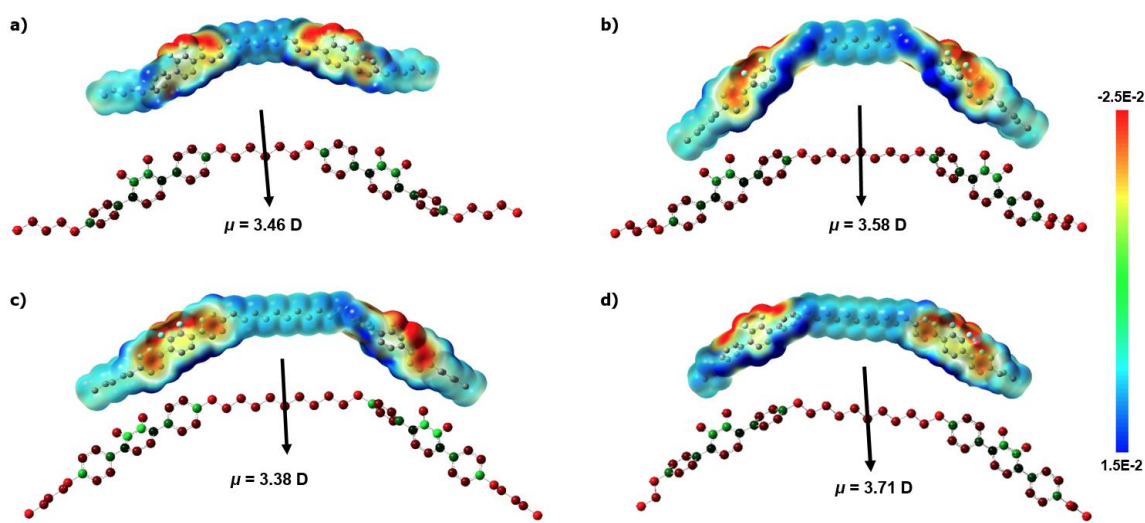

**Figure 6.** The electrostatic potential isosurfaces and ball-and-stick models showing the molecular dipole moment $\mathbf{\mu}$ of a) DTC5C7, b) DTC5C9, c) DTC5C11, d) DTC3C11. The red color indicates the electron-rich regions.

*3.2 Phase diagrams*

The phase diagrams of the fluorinated dimeric mesogens (Fig.1) that are presented in Table 1 are determined upon cooling at the rate of 0.1 °/min using polarizing optical microscopy (POM) observations. The temperature is controlled with the Instec HCS402 hot stage and mK2000 temperature controller (Instec, Inc.) with temperature stability of 0.01 ℃. Since the transitions are of the first order, which implies the presence of biphasic regions, superheating and supercooling, we list the transition temperatures with an accuracy of 1 ℃. The N to $N_{TB}$ transition in homogeneous planar cells is accompanied by the appearance of periodic stripes and focal conic domains, Fig. 7a.

**Table 1.** Phase diagrams of fluorinated liquid crystal dimers. The phase diagram is determined on cooling at the rate of 0.1 °C/min from the isotropic phase.

| Compound | $m$ | $n$ | Phase diagram |
|---|---|---|---|
| DTC5C7 | 5 | 7 | $N_{TB} \xleftarrow{129\,°C} N \xleftarrow{158\,°C} Iso$ |
| DTC5C9 | 5 | 9 | $N_{TB} \xleftarrow{124\,°C} N \xleftarrow{162\,°C} Iso$ |
| DTC5C11 | 5 | 11 | $N_{TB} \xleftarrow{124\,°C} N \xleftarrow{167\,°C} Iso$ |
| DTC3C11 | 3 | 11 | $N_{TB} \xleftarrow{128\,°C} N \xleftarrow{180\,°C} Iso$ |

*3.3 Liquid crystal alignment*

The planar alignment of studied dimers is realized via conventional alignment methods using polyimides [61]. We spin-coat PI2555 polyimide (HD Microsystems) on indium tin oxide (ITO)-coated glass substrates to achieve homogeneous planar alignment. The glass substrates are unidirectionally rubbed with a velvet cloth ten times and assembled in an antiparallel fashion. The cell gap, $d$, is controlled by Micropearl spherical spacers mixed with NOA71 UV glue and measured using a PerkinElmer UV/Vis Spectrometer Lambda 18.

Attempted homeotropic alignment by conventional polyimides or surfactants alone resulted either in a misaligned Schlieren texture or in a weak homeotropic alignment stable within a narrow temperature range, as upon cooling, the N phase experiences an anchoring transition [62], and the homeotropic alignment is lost [61]. To realize a stable homeotropic alignment of fluorinated dimers, we developed the following procedure. First, the ITO glass is cleaned in the ultrasonic bath at 60 ℃ for 30 minutes, rinsed in deionized (DI) water for three minutes, and rinsed with isopropyl alcohol. The substrates are placed in an oven preheated to $T = 90$ ℃ to evaporate the solvent. After drying, the ITO glass is treated with UV ozone for 15 minutes. Subsequently, the substrates are immersed and agitated in 1 wt% aqueous solution of Dimethyloctadecyl[3-(trimethoxysilyl)propyl]ammonium chloride (DMOAP) (Sigma-Aldrich, Cat. # 435694) for 25 min. The substrates are then rinsed with DI water for three minutes, dried with nitrogen gas, and cured in an oven at $T = 110$ °C. Lastly, the second alignment layer, SE5661 mixed with a thinner, Solvent 79, in a 1:1 ratio (Nissan Chemical Industries), is spin-coated onto the ITO-glass substrates in three consecutive steps at 500 rpm (3 sec), 3000 rpm (30 sec), 50 rpm (1 sec) without interruptions. After spin-coating, the substrates are soft-baked at $T = 80$ °C for 10 minutes, and, finally, baked at $T = 180$ ℃ for 55 minutes. Homeotropic alignment and its thermal stability are confirmed by conoscopy, Fig. 7b.

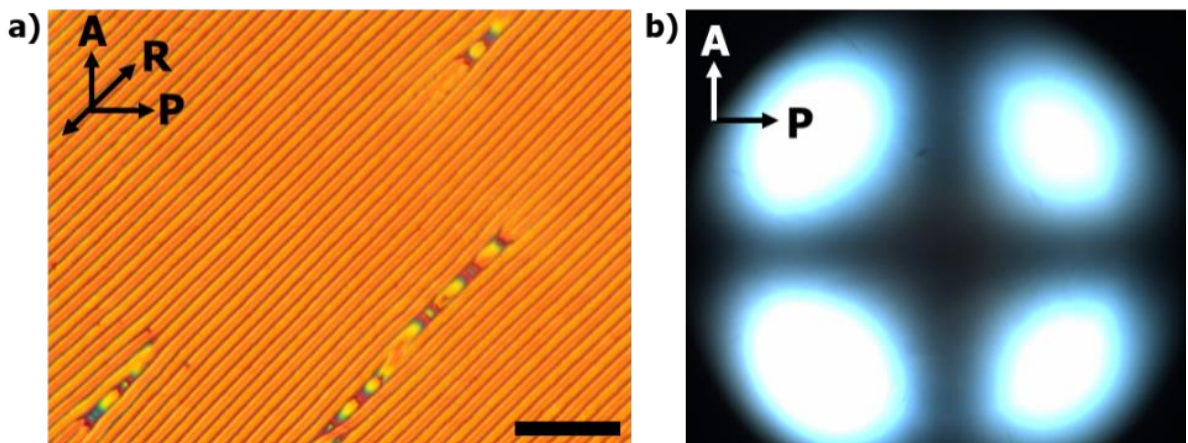

**Figure 7.** (a) Polarizing optical microscopy textures of N_TB phase of DTC3C11 in a planar cell ($d = 12.5$ μm, $T = 110$ ℃), and (b) conoscopic image of a homeotropic cell ($d = 15.5$ μm) in the N

phase ($T = 135.5$ °C). R indicates the rubbing direction, polarizer and analyzer are labeled as P and A, respectively. Scale bar in part (a) is 50 μm.

### *3.4 Freeze-fracture transmission electron microscopy of the $N_{TB}$ phase*

To preserve the intrinsic liquid crystal nanostructure, we prepare replica samples for FFTEM using plunge freezing [24]. First, the material is sandwiched between two copper planchettes and heated to the isotropic phase. Each sample is cooled down and stabilized at $T = 110$ °C. The planchettes are taken with tweezers and rapidly cooled in liquid ethane, followed by freezing in liquid nitrogen with an approximate cooling rate of $> 10^3$ °/min to prevent possible phase transitions. Next, the samples are transferred into a freeze-fracture vacuum chamber (BalTec BAF060) and kept at $T = -140$ °C. The built-in microtome is used to knock the top planchette and expose the frozen liquid crystal fractured surface. The fractured surface is consecutively shadowed and replicated by depositing ~4 nm Pt/C at 45 ° angle, and ~20 nm carbon from the top.

After removing the replicated samples from the freeze-fracture machine, the LC material is dissolved in chloroform. The replicas are collected onto Copper TEM 300 mesh grids with a lacey carbon coated support, (Pacific Grid-Tech). Finally, the replicas are observed using a TEM FEI Tecnai F20 at room temperature. The typical TEM structures demonstrating the nanoscale modulation of the molecular orientation are shown in Fig. 8a,b. The nanoscale pitch increases from about 8.1 nm to 9.6 nm as the spacer length increases, Fig. 8c.

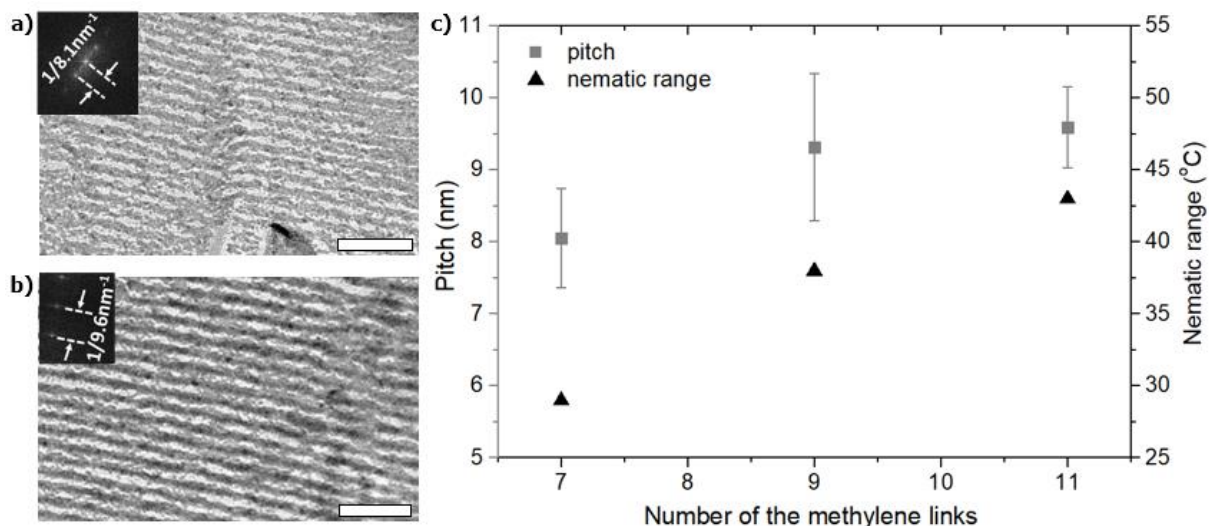

**Figure 8.** FFTEM images of (a) DTC5C7 and (b) DTC5C11 dimers in the $N_{TB}$ phase. The insets show the nanoscale pitch measured by taking a spatial Fourier transform of the periodic regions in the TEM images. Scale bars, 50 nm. (c) Nanoscale pitch and nematic range of DTC5C$n$ compounds, where $n$=7,9,11. The Pt/C replicas capture the $N_{TB}$ phase at 19 °C (DTC5C7) in (a) and at 14 °C (DTC5C9, DTC5C11) in (b) below $T_{NTB}$.

### *3.5 Birefringence*

The temperature dependence of the birefringence, $\Delta n(T)$, in thin planar cells (4.8 µm - 5.6 µm) is determined by measuring the optical retardance, $\Gamma(T) = \Delta n(T)d$, on cooling employing LC-PolScope (Abrio Imaging Systems). All samples are probed with monochromatic illumination at the wavelength $\lambda = 546$ nm. Since the retardance limit of the LC-PolScope is 273 nm, we also measure the optical retardance values using a Berek quartz wedge compensator and a $\lambda = 546$ nm filter (Olympus Co.). Figure 9a demonstrates that $\Delta n$ increases when the number of the methyl groups in the flexible alkyl bridge increases from $n = 7$ to 9 and 11. For the fixed $n =11$, $\Delta n$ also increases when the length of the terminal chains increases from $m = 3$ to 5. Upon cooling, $\Delta n$ increases since the orientational order is enhanced. However, a few degrees before the N-$N_{TB}$ transition, $\Delta n$ saturates, as in the case of DTC5C7, or even slightly decreases, as in the other three homologs, Fig. 9b.

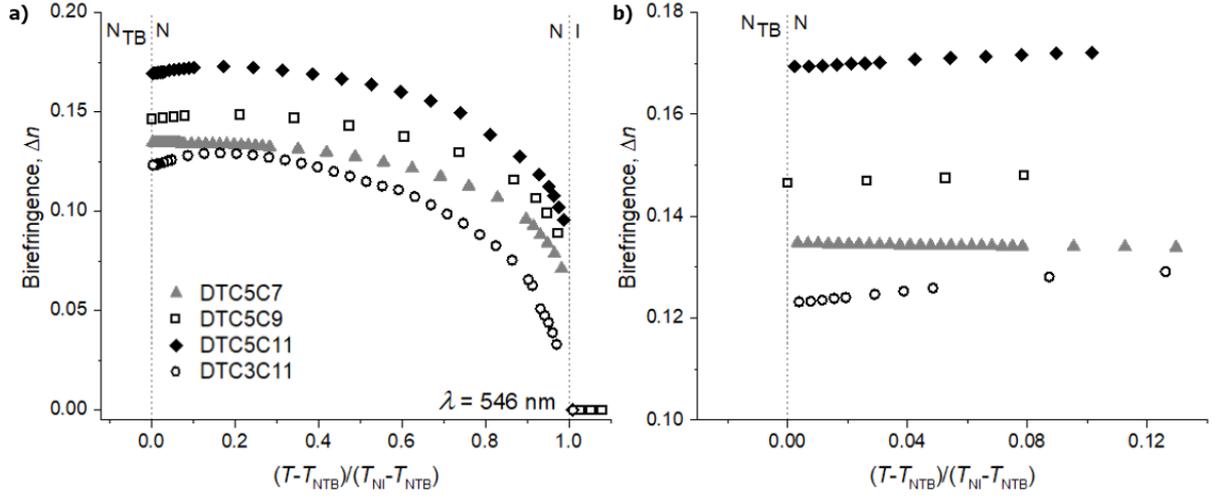

**Figure 9.** (a) Temperature dependence of $\Delta n(T)$ for DTC5C7 (gray triangles) DTC5C9 (open squares), DTC5C11 (filled diamonds), and DTC3C11 (open circles); (b) $\Delta n(T)$ near the N-N$_{TB}$ pretransitional region. The wavelength of the probing light is 546 nm. The error bars are smaller than the plot symbols.

### 3.6 Dielectric anisotropy

The dielectric properties are characterized by a two-cell method. The capacitance of LC cells with an active ITO electrode area of $5 \times 5 \text{ mm}^2$ is measured using an LCR meter HP4284A. Since dielectric permittivities are both temperature and frequency dependant, for each dimer, we first measure capacitance and resistance by applying $0.2 \text{ V}_{RMS}$ over a frequency range from 20 Hz to $10^6$ Hz. Further dielectric measurements for each material are carried out at frequencies at which the effects of conductivity and dielectric relaxation are minimum, namely, 20 kHz for DTC5C7, 40 kHz for DTC5C9, and 10 kHz for both DTC5C11 and DTC3C11; the Frederiks effects are explored at the same frequencies. The perpendicular component $\varepsilon_{\perp}$ of dielectric permittivity is measured in planar cells, while $\varepsilon_{\parallel}$ is determined in homeotropic cells. All temperature dependencies are measured on cooling. The dielectric anisotropy of all four homologs is negative, Fig. 10. The absolute value of $\Delta\varepsilon(T)$ increases when (i) the flexible alkyl bridges become longer and (ii) the terminal chains become shorter. Near $T_{NTB}$, $|\Delta\varepsilon(T)|$ tends to saturate or to slightly decrease.

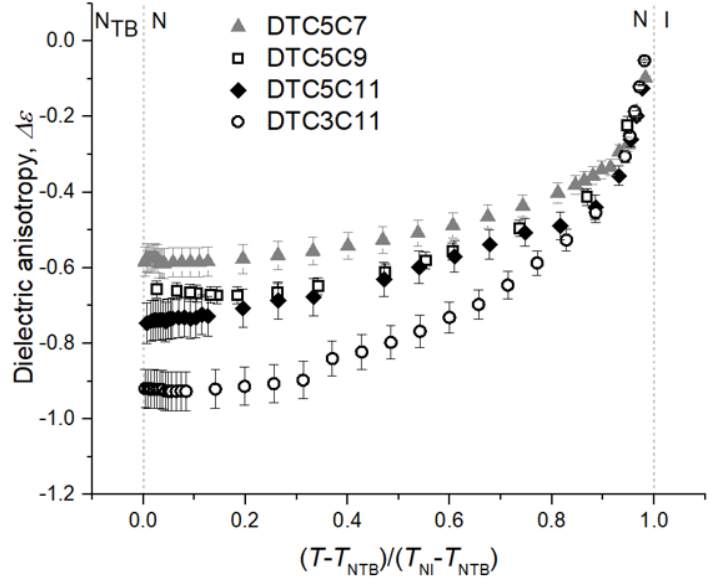

**Figure 10.** Temperature dependence of $\Delta\varepsilon(T)$ for liquid crystal dimers DTC5C7 (filled triangles), DTC5C9 (open squares), DTC5C11 (filled diamonds), and dimer DTC3C11 (open circles) measured at frequencies of 20, 40, 10, and 10 kHz, respectively.

### 3.7 Bend elastic constant

The bend elastic constant $K_{33}$ is determined by measuring the voltage threshold $V_{th}$ of the bend Frederiks transition in homeotropic cells with transparent electrodes at the bounding plates. We apply an AC electric field of the frequency specified above, parallel to the director, $\hat{\mathbf{n}}$, and measure the capacitance of the cells as a function of the applied voltage, using the LCR meter. $K_{33}$ is determined from the expression [63]

$$K_{33} = \frac{\varepsilon_o \Delta\varepsilon V_{th}^2}{\pi^2}.$$  (1)

The bend elastic modulus of all four compounds shows a strong nonmonotonous temperature dependence. Upon cooling from the isotropic phase, $K_{33}$ first sharply increases and then decreases on approaching $T_{NTB}$, to $\sim 0.5$ pN; very close to the N-$N_{TB}$ transition, $K_{33}$ increases slightly, Fig. 11. The maximum value of bend elastic modulus, $K_{33}$, increases with increasing the bridge length (compare DTC5C11 to DTC5C9 and DTC5C7 in Fig.11). Furthermore, comparison of DTC3C11 and DTC5C11 that differ only in the length of the terminal chains show that shorter terminal chains lead to a higher $K_{33}$; in fact, DTC3C11 shows the highest $K_{33}$ among the four studied substances.

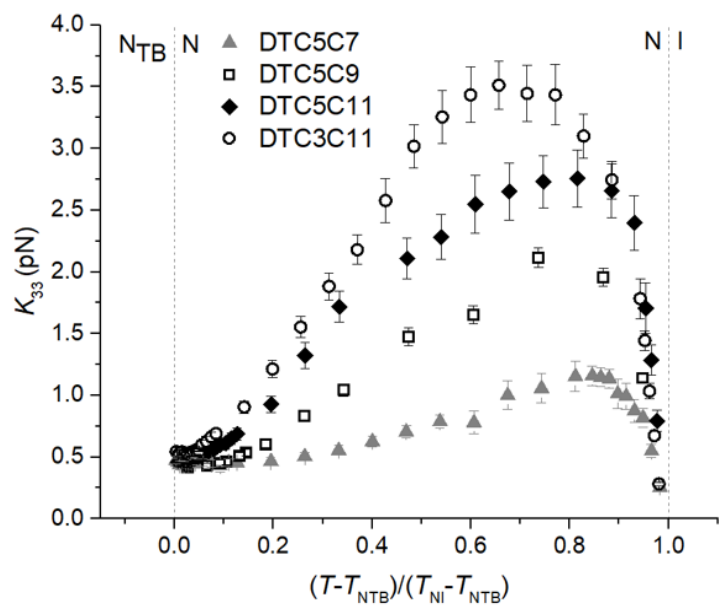

**Figure 11.** Temperature-dependent bend elastic modulus, $K_{33}$, of DTC5C7 (filled triangles), DTC5C9 (open squares), DTC5C11 (filled diamonds), and DTC3C11 (open circles) dimers.

## 4. Discussion and conclusion

In this report, we present the experimentally determined material parameters of four symmetric flexible dimeric homologs comprised of the 2',3'-difluoroterphenyls units that differ in the linking chain length as well as the length of the terminal moiety. All four exhibit the low-temperature $N_{TB}$ phase confirmed by FFTEM (Fig. 8), with the period that slightly increases when the flexible bridge connecting two 2',3'-difluoroterphenyl rigid units becomes longer. The result is consistent with the assumption that the molecular bend angle [64] and the helical pitch increase with the bridge length [65]. Previous reports reveal a similar correlation [65, 66]. The temperature range over which the N phase exists also increases with an increased bridge length, similarly to the well-studied CB$n$CB series with $n$=7,9,11 [67].

The rest of the material properties, namely, $\Delta n(T)$, $\Delta \varepsilon(T)$ and $K_{33}$, are measured in the high-temperature uniaxial N phase. Longer bridges and terminal chains yield a higher $\Delta n$, in agreement with the idea that longer molecules show higher polarizability [68]. Additionally, $\Delta n(T)$ either saturates or exhibits a slight pretransitional decrease near $T_{\text{NTB}}$ (Fig. 9b). Other $N_{\text{TB}}$-forming flexible dimers show a similar pretransitional behavior of $\Delta n(T)$ [13, 40, 41, 45, 46, 69].

Each dimer molecule possesses two strong permanent dipole moments associated with the C-F bonds. The net dipole for all four mesogens is perpendicular to the long axis, yielding a negative dielectric anisotropy, $\Delta \varepsilon < 0$ (Figs. 6,10). Increase in the bridge length or decrease in the length of the terminal moiety increases $|\Delta \varepsilon(T)|$, a behavior associated with a complex combination of permanent dipoles, polarizability, and molecular conformations. $N_{\text{TB}}$-forming flexible dimers might adopt stronger bend conformations near $T_{\text{NTB}}$ [19, 70, 71, 72, 73], which might explain a slight pretransitional decrease in $|\Delta \varepsilon(T)|$ near the N-$N_{\text{TB}}$ transition, Fig. 10.

A novel technique of vertical alignment of the fluorinated dimers is developed to directly measure the bend elastic modulus $K_{33}$, using the electric Frederiks transition. The measured $K_{33}$ shows a non-monotonous temperature dependency, Fig. 11. Upon cooling from the isotropic phase, $K_{33}$ first increases, similarly to other $N_{\text{TB}}$ compounds [39, 41, 42, 44, 45, 46, 74] and conventional nematics formed by rod-like LC molecules. After reaching a maximum, $K_{33}$ of all four dimes starts to decrease, approaching $\sim 0.5$ pN near the N-$N_{\text{TB}}$ transition.

The obtained results show that even small changes in the length of a flexible aliphatic bridge of the $N_{\text{TB}}$-forming fluorinated dimers yield noticeable modifications of the material parameters. The data might be of interest in formulating mixtures with predesigned properties.

**Acknowledgments**

The freeze-fracture transmission electron microscopy studies were performed at the (cryo) TEM facility at the Advanced Materials and Liquid Crystal Institute, Kent State University, supported by the Ohio Research Scholars Program Research Cluster on Surfaces in Advanced

Materials.

**Disclosure statement**

No potential conflict of interest was reported by the authors.

**Funding**

The work was supported by the NSF grant ECCS-1906104.

**References**

1. Gleeson HF, Kaur S, Görtz V, Belaissaoui A, Cowling S, Goodby JW. The nematic phases of bent-core liquid crystals. ChemPhysChem. 2014;15:1251-60.
2. Jákli A, Lavrentovich OD, Selinger JV. Physics of liquid crystals of bent-shaped molecules. Rev Mod Phys. 2018;90:045004.
3. Sadashiva BK, Reddy RA, Pratibha R, Madhusudana NV. Biaxial smectic A phase in homologous series of compounds composed of highly polar unsymmetrically substituted bent-core molecules. J Mater Chem. 2002;12:943-50.
4. Murthy HNS, Sadashiva BK. A polar biaxial smectic A phase in new unsymmetrical compounds composed of bent-core molecules. Liq Cryst. 2004;31:567-78.
5. Reddy RA, Sadashiva BK. Direct transition from a nematic phase to a polar biaxial smectic A phase in a homologous series of unsymmetrically substituted bent-core compounds. J Mater Chem. 2004;14:310-9.
6. Pratibha R, Madhusudana NV, Sadashiva BK. An orientational transition of bent-core molecules in an anisotropic matrix. Science. 2000;288:2184-7.
7. Smalyukh II, Pratibha R, Madhusudana NV, Lavrentovich OD. Selective imaging of 3D director fields and study of defects in biaxial smectic A liquid crystals. Eur Phys J E. 2005;16:179-91.
8. Sathyanarayana P, Mathew M, Li Q, Sastry VSS, Kundu B, Le KV, Takezoe H, Dhara S. Splay bend elasticity of a bent-core nematic liquid crystal. Phys Rev E. 2010;81:010702.
9. Majumdar M, Salamon P, Jákli A, Gleeson JT, Sprunt S. Elastic constants and orientational viscosities of a bent-core nematic liquid crystal. Phys Rev E. 2011;83:031701.
10. Kaur S, Liu H, Addis J, Greco C, Ferrarini A, Gortz V, Goodby JW, Gleeson HF. The influence of structure on the elastic, optical and dielectric properties of nematic phases formed from bent-core molecules. J Mater Chem C. 2013;1:6667-76.
11. Sathyanarayana P, Radhika S, Sadashiva BK, Dhara S. Structure–property correlation of a hockey stick-shaped compound exhibiting N-SmA-SmCa phase transitions. Soft Matter. 2012;8:2322-7.
12. Avci N, Borshch V, Sarkar DD, Deb R, Venkatesh G, Turiv T, Shiyanovskii SV, Rao NVS, Lavrentovich OD. Viscoelasticity, dielectric anisotropy, and birefringence in the nematic phase of three four-ring bent-core liquid crystals with an L-shaped molecular frame. Soft Matter. 2013;9:1066-75.
13. Cukrov G, Golestani YM, Xiang J, Nastishin YA, Ahmed Z, Welch C, Mehl GH, Lavrentovich OD. Comparative analysis of anisotropic material properties of uniaxial nematics formed by flexible dimers and rod-like monomers. Liq Cryst. 2017;44:219-31.
14. Kleman M, Lavrentovich OD. Soft Matter Physics: An Introduction. New York: Springer; 2003.
15. Li BX, Nastishin YA, Wang H, Gao M, Paladugu S, Li R, Fukuto M, Li Q, Shiyanovskii SV, Lavrentovich OD. Liquid crystal phases with unusual structures and physical properties formed by acute-angle bent core molecules. Physical Review Research. 2020;2:033371.
16. Lee JH, Yoon TH, Choi EJ. Unusual temperature dependence of the splay elastic constant of a rodlike nematic liquid crystal doped with a highly kinked bent-core molecule. Phys Rev E. 2013;88:062511.

17.     Meyer RB. Structural Problems in Liquid Crystal Physics. In: Balian R, Weill G, editors. Molecular Fluids Les Houches Lectures, 1973. Les Houches Gordon and Breach; 1976. p. 271-343.

18.     Dozov I. On the spontaneous symmetry breaking in the mesophases of achiral banana-shaped molecules. Europhys Lett. 2001;56:247-53.

19.     Cestari M, Diez-Berart S, Dunmur DA, Ferrarini A, de la Fuente MR, Jackson DJB, Lopez DO, Luckhurst GR, Perez-Jubindo MA, Richardson RM, Salud J, Timimi BA, Zimmermann H. Phase behavior and properties of the liquid-crystal dimer 1 '',7 ''-bis(4-cyanobiphenyl-4 '- yl) heptane: A twist-bend nematic liquid crystal. Phys Rev E. 2011;84.

20.     Chen D, Porada JH, Hooper JB, Klittnick A, Shen YQ, Tuchband MR, Korblova E, Bedrov D, Walba DM, Glaser MA, Maclennan JE, Clark NA. Chiral heliconical ground state of nanoscale pitch in a nematic liquid crystal of achiral molecular dimers. P Natl Acad Sci USA. 2013;110:15931-6.

21.     Borshch V, Kim YK, Xiang J, Gao M, Jakli A, Panov VP, Vij JK, Imrie CT, Tamba MG, Mehl GH, Lavrentovich OD. Nematic twist-bend phase with nanoscale modulation of molecular orientation. Nat Commun. 2013;4.

22.     Xiang J, Shiyanovskii SV, Imrie CT, Lavrentovich OD. Electrooptic Response of Chiral Nematic Liquid Crystals with Oblique Helicoidal Director. Phys Rev Lett. 2014;112:217801.

23.     Iadlovska OS, Babakhanova G, Mehl GH, Welch C, Cruickshank E, Strachan GJ, Storey JMD, Imrie CT, Shiyanovskii SV, Lavrentovich OD. Temperature dependence of bend elastic constant in oblique helicoidal cholesterics. Physical Review Research. 2020;2:013248.

24.     Gao M, Kim YK, Zhang CY, Borshch V, Zhou S, Park HS, Jakli A, Lavrentovich OD, Tamba MG, Kohlmeier A, Mehl GH, Weissflog W, Studer D, Zuber B, Gnagi H, Lin F. Direct Observation of Liquid Crystals Using Cryo-TEM: Specimen Preparation and Low-Dose Imaging. Microscopy Research and Technique. 2014;77:754-72.

25.     Meyer RB. Effects of Electric and Magnetic Fields on Structure of Cholesteric Liquid Crystals. Appl Phys Lett. 1968;12:281-2.

26.     Chen D, Nakata M, Shao R, Tuchband MR, Shuai M, Baumeister U, Weissflog W, Walba DM, Glaser MA, Maclennan JE, Clark NA. Twist-bend heliconical chiral nematic liquid crystal phase of an achiral rigid bent-core mesogen. Phys Rev E Stat Nonlin Soft Matter Phys. 2014;89:022506.

27.     Gorecka E, Salamonczyk M, Zep A, Pociecha D, Welch C, Ahmed Z, Mehl GH. Do the short helices exist in the nematic TB phase? Liq Cryst. 2015;42:1-7.

28.     Tuchband MR, Shuai M, Graber KA, Chen D, Radzihovsky L, Klittnick A, Foley L, Scarbrough A, Porada JH, Moran M. The twist-bend nematic phase of bent mesogenic dimer CB7CB and its mixtures. arXiv preprint arXiv:151107523. 2015.

29.     Paterson DA, Gao M, Kim YK, Jamali A, Finley KL, Robles-Hernandez B, Diez-Berart S, Salud J, de la Fuente MR, Timimi BA, Zimmermann H, Greco C, Ferrarini A, Storey JMD, Lopez DO, Lavrentovich OD, Luckhurst GR, Imrie CT. Understanding the twist-bend nematic phase: the characterisation of 1-(4-cyanobiphenyl-4 '-yloxy)-6-(4-cyanobiphenyl-4 '-yl)hexane (CB6OCB) and comparison with CB7CB. Soft Matter. 2016;12:6827-40.

30.     Paterson DA, Xiang J, Singh G, Walker R, Agra-Kooijman DM, Martinez-Felipe A, Gao M, Storey JM, Kumar S, Lavrentovich OD, Imrie CT. Reversible Isothermal Twist-Bend Nematic-Nematic Phase Transition Driven by the Photoisomerization of an Azobenzene-Based Nonsymmetric Liquid Crystal Dimer. J Am Chem Soc. 2016;138:5283-9.

31.     Zhu C, Tuchband MR, Young A, Shuai M, Scarbrough A, Walba DM, Maclennan JE, Wang C, Hexemer A, Clark NA. Resonant Carbon K-Edge Soft X-Ray Scattering from Lattice-Free Heliconical Molecular Ordering: Soft Dilative Elasticity of the Twist-Bend Liquid Crystal Phase. Phys Rev Lett. 2016;116:147803.

32.     Shamid SM, Dhakal S, Selinger JV. Statistical mechanics of bend flexoelectricity and the twist-bend phase in bent-core liquid crystals. Phys Rev E Stat Nonlin Soft Matter Phys. 2013;87:052503.

33.     Singh S. Curvature elasticity in liquid crystals. Phys Rep. 1996;277:284-384.

34.     Xiang J, Li YN, Li Q, Paterson DA, Storey JMD, Imrie CT, Lavrentovich OD. Electrically Tunable Selective Reflection of Light from Ultraviolet to Visible and Infrared by Heliconical Cholesterics. Adv Mater. 2015;27:3014-8.

35.     Salili SM, Xiang J, Wang H, Li Q, Paterson DA, Storey JMD, Imrie CT, Lavrentovich OD, Sprunt SN, Gleeson JT, Jakli A. Magnetically tunable selective reflection of light by heliconical cholesterics. Phys Rev E. 2016;94:042705.

36.     Gennes PGd. Calcul de la distorsion d'une structure cholesterique par un champ magnetique. Solid State Commun. 1968;6:163-5.

37.     Adlem K, Copic M, Luckhurst GR, Mertelj A, Parri O, Richardson RM, Snow BD, Timimi BA, Tuffin RP, Wilkes D. Chemically induced twist-bend nematic liquid crystals, liquid crystal dimers, and negative elastic constants. Physical review E, Statistical, nonlinear, and soft matter physics. 2013;88:022503.

38.     Yun CJ, Vengatesan MR, Vij JK, Song JK. Hierarchical elasticity of bimesogenic liquid crystals with twist-bend nematic phase. Appl Phys Lett. 2015;106:173102.

39.     Lopez DO, Robles-Hernandez B, Salud J, de la Fuente MR, Sebastian N, Diez-Berart S, Jaen X, Dunmur DA, Luckhurst GR. Miscibility studies of two twist-bend nematic liquid crystal dimers with different average molecular curvatures. A comparison between experimental data and predictions of a Landau mean-field theory for the N-TB-N phase transition. Phys Chem Chem Phys. 2016;18:4394-404.

40.     Sebastián N, Tamba MG, Stannarius R, de la Fuente MR, Salamonczyk M, Cukrov G, Gleeson J, Sprunt S, Jákli A, Welch C, Ahmed Z, Mehl GH, Eremin A. Mesophase structure and behaviour in bulk and restricted geometry of a dimeric compound exhibiting a nematic-nematic transition. Phys Chem Chem Phys. 2016;18:19299-308.

41.     Babakhanova G, Parsouzi Z, Paladugu S, Wang H, Nastishin YA, Shiyanovskii SV, Sprunt S, Lavrentovich OD. Elastic and viscous properties of the nematic dimer CB7CB. Phys Rev E. 2017;96:062704.

42.     Parsouzi Z, Babakhanova G, Rajabi M, Saha R, Gyawali P, Turiv T, Wang H, Baldwin AR, Welch C, Mehl GH, Gleeson JT, Jakli A, Lavrentovich OD, Sprunt S. Pretransitional behavior of viscoelastic parameters at the nematic to twist-bend nematic phase transition in flexible n-mers. Phys Chem Chem Phys. 2019;21:13078-89.

43.     Robles-Hernandez B, Sebastian N, de la Fuente MR, Lopez DO, Diez-Berart S, Salud J, Ros MB, Dunmur DA, Luckhurst GR, Timimi BA. Twist, tilt, and orientational order at the nematic to twist-bend nematic phase transition of 1'',9''-bis(4-cyanobiphenyl-4'-yl) nonane: A dielectric, (2)H NMR, and calorimetric study. Phys Rev E Stat Nonlin Soft Matter Phys. 2015;92:062505.

44.     Sebastián N, Robles-Hernandezb B, Diez-Berart S, Saludc J, Luckhurst GR, Dunmur DA, Lopez DO, La Fuente MR. Distinctive dielectric properties of nematic liquid crystal dimers. Liq Cryst. 2017;44:177-90.

45.     Saha R, Babakhanova G, Parsouzi Z, Rajabi M, Gyawali P, Welch C, Mehl GH, Gleeson J, Lavrentovich OD, Sprunt S, Jakli A. Oligomeric odd-even effect in liquid crystals. Mater Horizons. 2019;6:1905-12.

46.     Pociecha D, Crawford CA, Paterson DA, Storey JMD, Imrie CT, Vaupotic N, Gorecka E. Critical behavior of the optical birefringence at the nematic to twist-bend nematic phase transition. Phys Rev E. 2018;98:052706.

47.     Imrie CT. Liquid Crystal Dimers. In: Mingos DMP, editor. Liquid Crystals II. Berlin, Heidelberg: Springer Berlin Heidelberg; 1999. p. 149-92.

48.     Mandle RJ. Designing Liquid-Crystalline Oligomers to Exhibit Twist-Bend Modulated Nematic Phases. Chem Rec. 2018;18:1341-9.

49.     Abberley JP, Storey JMD, Imrie CT. Structure-property relationships in azobenzene-based twist-bend nematogens. Liq Cryst. 2019;46:2102-14.

50.     Forsyth E, Paterson DA, Cruickshank E, Strachan GJ, Gorecka E, Walker R, Storey JMD, Imrie CT. Liquid crystal dimers and the twist-bend nematic phase: On the role of spacers and terminal alkyl chains. J Mol Liq. 2020;320.

51.     Henderson PA, Imrie CT. Methylene-linked liquid crystal dimers and the twist-bend nematic phase. Liq Cryst. 2011;38:1407-14.

52.     Mandle RJ, Davis EJ, Voll CCA, Archbold CT, Goodby JW, Cowling SJ. The relationship between molecular structure and the incidence of the N-TB phase. Liq Cryst. 2015;42:688-703.

53.     Mandle RJ, Davis EJ, Archbold CT, Voll CCA, Andrews JL, Cowling SJ, Goodby JW. Apolar Bimesogens and the Incidence of the Twist-Bend Nematic Phase. Chem-Eur J. 2015;21:8158-67.

54.     Mandle RJ. The dependency of twist-bend nematic liquid crystals on molecular structure: a progression from dimers to trimers, oligomers and polymers. Soft Matter. 2016;12:7883-901.

55.     Ivsic T, Vinkovic M, Baumeister U, Mikleusevic A, Lesac A. Towards understanding the N-TB phase: a combined experimental, computational and spectroscopic study. Rsc Adv. 2016;6:5000-7.

56.     Dawood AA, Grossel MC, Luckhurst GR, Richardson RM, Timimi BA, Wells NJ, Yousif YZ. Twist-bend nematics, liquid crystal dimers, structure-property relations. Liq Cryst. 2017;44:106-26.

57.     Abberley JP, Jansze SM, Walker R, Paterson DA, Henderson PA, Marcelis ATM, Storey JMD, Imrie CT. Structure-property relationships in twist-bend nematogens: the influence of terminal groups. Liq Cryst. 2017;44:68-83.

58.     Stevenson WD, Zou HX, Zeng XB, Welch C, Ungar G, Mehl GH. Dynamic calorimetry and XRD studies of the nematic and twist-bend nematic phase transitions in a series of dimers with increasing spacer length. Phys Chem Chem Phys. 2018;20:25268-74.

59.     Ziauddin Z. The design, synthesis and properties of dimeric molecules exhibiting nematic-nematic transitions: The University of Hull; 2014.

60.     Ahmed Z, Welch C, Mehl GH. The design and investigation of the self-assembly of dimers with two nematic phases. Rsc Adv. 2015;5:93513-21.

61.     Babakhanova G, Lavrentovich OD. The Techniques of Surface Alignment of Liquid Crystals: Springer International Publishing; 2019.

62.     Kim YK, Cukrov G, Xiang J, Shin ST, Lavrentovich OD. Domain walls and anchoring transitions mimicking nematic biaxiality in the oxadiazole bent-core liquid crystal C7. Soft Matter. 2015;11:3963-70.

63.     Blinov LM, Chigrinov VG. Electrooptic effects in liquid crystal materials. New York: Springer-Verlag; 1994.

64.     Greco C, Luckhurst GR, Ferrarini A. Molecular geometry, twist-bend nematic phase and unconventional elasticity: a generalised Maier-Saupe theory. Soft Matter. 2014;10:9318-23.

65.     Saha R, Feng C, Welch C, Mehl GH, Feng J, Zhu C, Gleeson J, Sprunt S, Jákli A. The interplay between spatial and heliconical orientational order in twist-bend nematic materials. Physical Chemistry Chemical Physics. 2021;23:4055-63.

66.     Arakawa Y, Komatsu K, Feng J, Zhu C, Tsuji H. Distinct twist-bend nematic phase behaviors associated with the ester-linkage direction of thioether-linked liquid crystal dimers. Materials Advances. 2021;2:261-72.

67.     Paterson DA, Abberley JP, Harrison WT, Storey JM, Imrie CT. Cyanobiphenyl-based liquid crystal dimers and the twist-bend nematic phase. Liq Cryst. 2017;44:127-46.

68.     Collings PJ, Hird M. Introduction to liquid crystals chemistry and physics. London ; Bristol, PA: Taylor & Francis; 1997.

69.     Meyer C, Luckhurst GR, Dozov I. The temperature dependence of the heliconical tilt angle in the twist-bend nematic phase of the odd dimer CB7CB. J Mater Chem C. 2015;3:318-28.

70.     Ferrarini A, Luckhurst GR, Nordio PL. Even-Odd Effects in Liquid-Crystal Dimers with Flexible Spacers - a Test of the Rotational Isomeric State Approximation. Mol Phys. 1995;85:131-43.

71.     Ferrarini A, Luckhurst GR, Nordio PL, Roskilly SJ. Understanding the Unusual Transitional Behavior of Liquid-Crystal Dimers. Chem Phys Lett. 1993;214:409-17.

72.     Ferrarini A, Luckhurst GR, Nordio PL, Roskilly SJ. Prediction of the Transitional Properties of Liquid-Crystal Dimers - a Molecular-Field Calculation Based on the Surface Tensor Parametrization. J Chem Phys. 1994;100:1460-9.

73.     Ferrarini A, Luckhurst GR, Nordio PL, Roskilly SJ. Understanding the dependence of the transitional properties of liquid crystal dimers on their molecular geometry. Liq Cryst. 1996;21:373-82.

74.     Balachandran R, Panov VP, Vij JK, Kocot A, Tamba MG, Kohlmeier A, Mehl GH. Elastic properties of bimesogenic liquid crystals. Liq Cryst. 2013;40:681-8.